\documentclass{vgtc}                          

\graphicspath{{figures/}{pictures/}{images/}{./}} 

\usepackage{times}                     

\usepackage{tabu}                      
\usepackage{booktabs}                  
\usepackage{lipsum}                    
\usepackage{mwe}                       

\usepackage{mathptmx}                  

\onlineid{0}

\vgtccategory{Research}

\vgtcinsertpkg

\title{Zoomable Level-of-Detail ChartTables for Interpreting Probabilistic Model Outputs for Reactionary Train Delays}

\author{Aidan Slingsby\thanks{e-mail: a.slingsby@city.ac.uk}\\ %
        \scriptsize City, University of London %
\and Jonathan Hyde\thanks{e-mail: jonathan.hyde@risksol.co.uk}\\ %
     \scriptsize Risk Solutions%
}

\teaser{
 \centering
  \includegraphics[width=\linewidth]{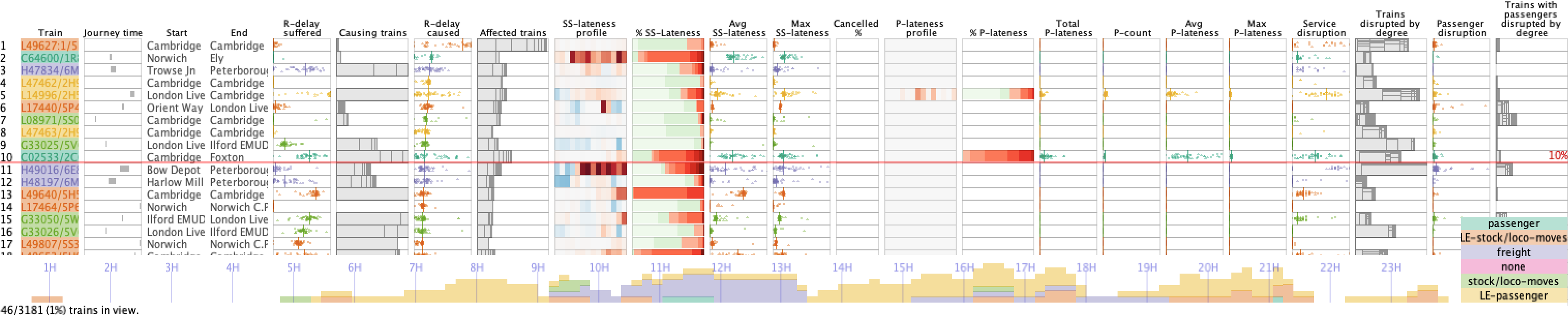}
  \caption{A Zoomable Level-of-Detail ChartTable, in which train delay metrics (columns) are represented as mini-charts for each train (row). The time histogram at the bottom indicates the number of trains operating over a 24h period and can be used to interactively filter trains by hour of day.}
  \label{fig:teaser}
}

\abstract{``Reactionary delay'' is a result of the accumulated cascading effects of knock-on train delays which is increasing on UK railways due to increasing utilisation of the railway infrastructure. The chaotic nature of its effects on train lateness is notoriously hard to predict. We use a stochastic Monte-Carto-style simulation of reactionary delay that produces whole distributions of likely reactionary delay and delays this causes. We demonstrate how Zoomable Level-of-Detail ChartTables -- case-by-variable tables where cases are rows, variables are columns, variables are complex composite metrics that incorporate distributions, and cells contain \textit{mini-charts} that depict these as different \textit{levels of detail} through zoom interaction -- help interpret whole \textit{distributions} of model outputs to help understand the causes and effects of reactionary delay, how they inform timetable robustness testing, and how they could be used in other contexts.} 


\usepackage{calc}
\usepackage{enumitem}

\begin{document}


\firstsection{Introduction}

\maketitle

We demonstrate how Zoomable Level-of-Detail ChartTables can help interpret probabilistic Monte-Carto-style simulations of reactionary delay and its effect on train lateness. ``Reactionary delay'' is the result of the accumulated cascading effects of knock-on train delays \cite{reactionary}. Its interdependent ``knock-on'' nature makes its effects hard to predict, with resulting delays often longer than the original delays. Reactionary delay is an increasing problem on UK railways as the number of scheduled train services are increasing \cite{rssb}. Our approach to helping understand the impact of reactionary delay is to use a stochastic ``Monte-Carlo'' style Agent-Based Model (\href{https://rs-rpm-demo.azurewebsites.net/}{``SaviRPM''}) that simulates trains running to a fixed timetable. Each model run incorporates randomly generated primary delays (e.g.\ late departure due to passenger overcrowding) based on historically-derived probabilities, where each run represents a possible alternative ``day'', detailing primary delays, resulting reactionary delays and their impacts on other trains. We can use this for comparing timetables for robustness and resilience.

Key to our approach is to enable consideration of the \textit{distribution} stochastic outputs using interactive visualisation. Agent-Based Models (ABMs; particularly Individual-Based Models) are often used to simulate individual behaviour to predict resulting larger-scale processes \cite{railsback2019agent}. We are modelling individual trains to help understand the accumulation of multiple knock-on effects at a system level. Understanding mechanisms within such models are so challenging that rich probabilistic data are \textit{often reduced to high-level summaries} \cite{Grainger2016}. Information Visualisation and Visual Analytics are often cited solutions to enable analysts to consider and interpret more detail and nuance than through high-level summaries alone.

\begin{figure*}[t!]
    \centering
    \includegraphics[width=\linewidth+7mm,trim=0px 0px 0px 0px, clip=true]{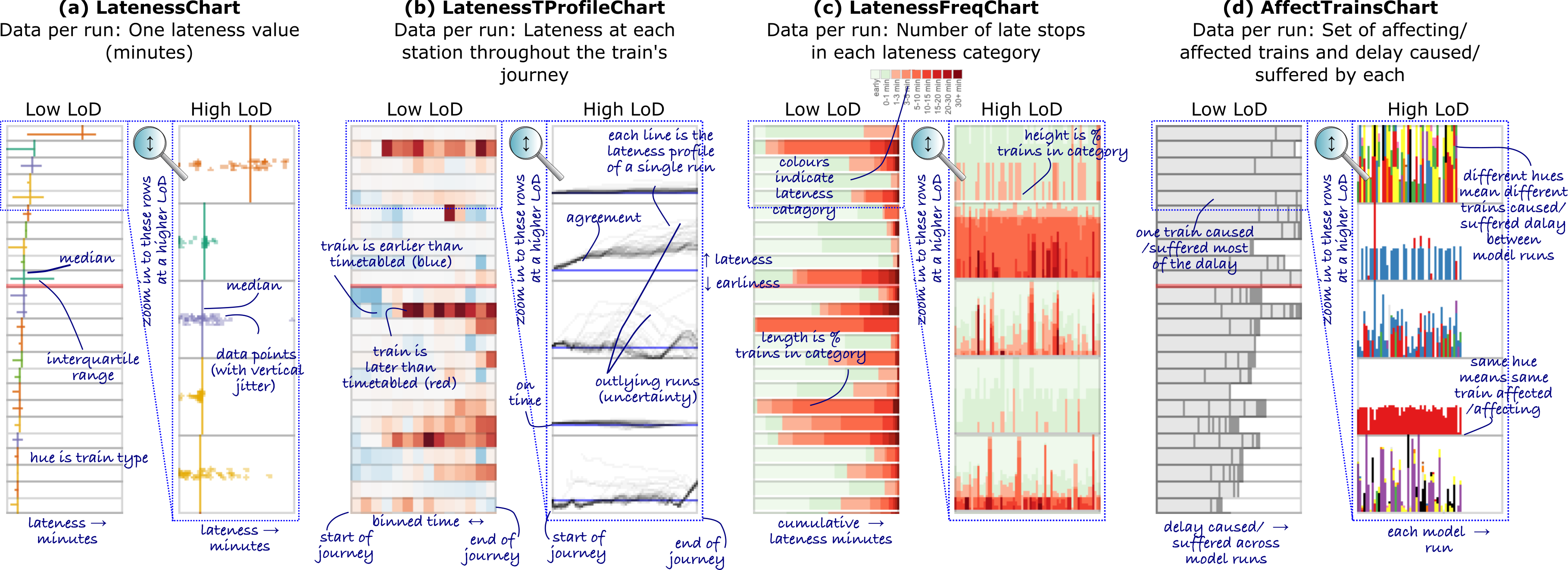}
    \caption{Four metric types depicted by four different mini-charts. Each has both low and high level-of-detail (LoD) variants. In all cases here, the high LoD variant on the right shows the detail of the top five rows of the low LoD on the left (outlined in a blue dotted line). An appropriate threshold then vertically zooming rows determines which variant is used.}
    \label{fig:types}
\end{figure*}

In our Zoomable Level-of-Detail ChartTables, rows are \textit{trains}, columns are \textit{composite metrics} that quantify aspects of reactionary delay, visually-represented as \textit{mini-charts} within cells that represent the metric for each train across all model runs. Four types of mini-chart (section \ref{sec:types}) for four different metric types have visual representations at two levels-of-detail depending on the zoom level (Fig.\ \ref{fig:types}). Vertical zooming changes the height of rows. Where rows are narrow and numerous, low level-of-detail mini-charts summarise across all trains. Where zoomed rows are wide/tall, high level-of-detail mini-charts depict the distribution of model runs.

Whilst many of these design characteristics are not new, we show how our design and interactions enable interpretation of probabilistic Monte-Carlo-style simulation for identifying the effects of reactionary delay. We show how they are helping the UK's railway industry understand reactionary delay and inform timetable redesign to reduce the effects of reactionary delay. We reflect more widely on how they can be used in other contexts to facilitate comparison and interpretation of complex metrics and their distributions.

Our contributions are to: \textit{(a)}~present Zoomable Level-of-Detail ChartTables, \textit{(b)}~describe a set of composite metrics and visual representations (mini-charts) that capture different aspects of probabilistic reactionary delay at different levels of detail, \textit{(c)}~demonstrate their use for interpreting stochastic Monte-Carlo-style simulation results,  and \textit{(d)}~reflect on their potential wider use.

\section{Zoomable Level-of-Detail ChartTables}

This work arose from a series of workshops and projects starting in 2018. A workshop run by \href{https://www.rssb.co.uk/}{Rail Safety \& Standards Board (RSSB)} (which funds applied research for the UK rail industry) established a need for a better understanding of reactionary delay to help improve punctuality. They published a call for applied research. We partnered with two UK Train Operating Companies (\href{https://www.gwr.com/}{Great Western Railways} and \href{https://www.greateranglia.co.uk/}{Greater Anglia}) and our research proposal -- to investigate the feasibility of combining stochastic Agent-Based Modelling with highly interactive visualisation -- was accepted and funded by RSSB. Our approach was shown to be feasible and further funded projects (to develop the ideas) and private consultancy (see acknowledgements; to apply and assess the feasibility, robustness and resilience of alternative timetables). Throughout, interactive visualisation was key for validating and interpreting the modelling. The work has been deeply embedded in industry needs. Early stages of the work used Koh \textit{et al}'s user-based approach of workshops, followed by iterative prototyping loosely based on the AGILE principles of short development/feedback cycles and continual re-prioritisation. This work now underpins consultancy and the software (\href{https://rs-rpm-demo.azurewebsites.net/}{``SaviRPM''}) as well as is being licensed to RSSB and a Train Operating Company for their use. In this short paper, we specifically focus on the Zoomable Level-of-Detail ChartTable technique from this work. 

We established some requirements to drive our designs:

\begin{itemize}[topsep=1em,parsep=0em,itemsep=0em]
    \item [R1:] \textbf{Depict multiple metrics}. Multiple aspects of lateness help us quantify delays, establish consequences, and identify problematic trains. Metrics include those that relate to delays caused by the train, delays suffered by the train, many of which cause train arrival lateness at stations and subsequent passenger arrival lateness (\href{https://osf.io/u2ykd/}{listed here}).
    
    \item [R2:] \textbf{Summarise composite metrics}. Although some metrics are single values, many are composite and cannot be expressed with a single number. See section \ref{sec:types} the metric types we used.
    
    \item [R3:] \textbf{Depict distribution of metrics across trains}. This is to help rank trains based on their delay metrics; i.e.\ their different contributions to delays or vulnerability to delay.
    
    \item [R4:] \textbf{Depict distribution of metrics across model runs}.  This is to help use consider the consistency of different types of delays between model runs and to help where there are plausibly likely serious worse cases. This aspect of probablistic modelling is often neglected, yet considered important.
\end{itemize}

For R1, we used the commonly-used case-by-variable table \cite{rao1994table} where cases (trains) are rows and variables (metrics) are columns, enabling multiple metrics to be considered (\href{https://osf.io/u2ykd/}{listed here}).

R2, R3 and R4 call for a solution that can depict complex metrics. Our solution is to use mini-charts that are embedded within cells. These depict composite metric values using different chart types for different metric types. Bertin demonstrated the value of depicting data using visual variables \cite{bertin1983semiology}. Mini-charts in tables are widely used for both single value metrics \cite{perin2014revisiting}, for multiple levels of abstraction \cite{beecham2016faceted} and for more complex metrics that summarise various types of aggregations \cite{furmanova2020taggle, li2024coinsight,li2022hitailor}. Mini-charts are also used in multivariate geographical mapping \cite{wickham2012glyph,slingsby2018tilemaps,slingsby2023gridded}. 

\begin{figure}
    \centering
    \includegraphics[width=\linewidth,trim=0px 100px 0px 0px, clip=true]{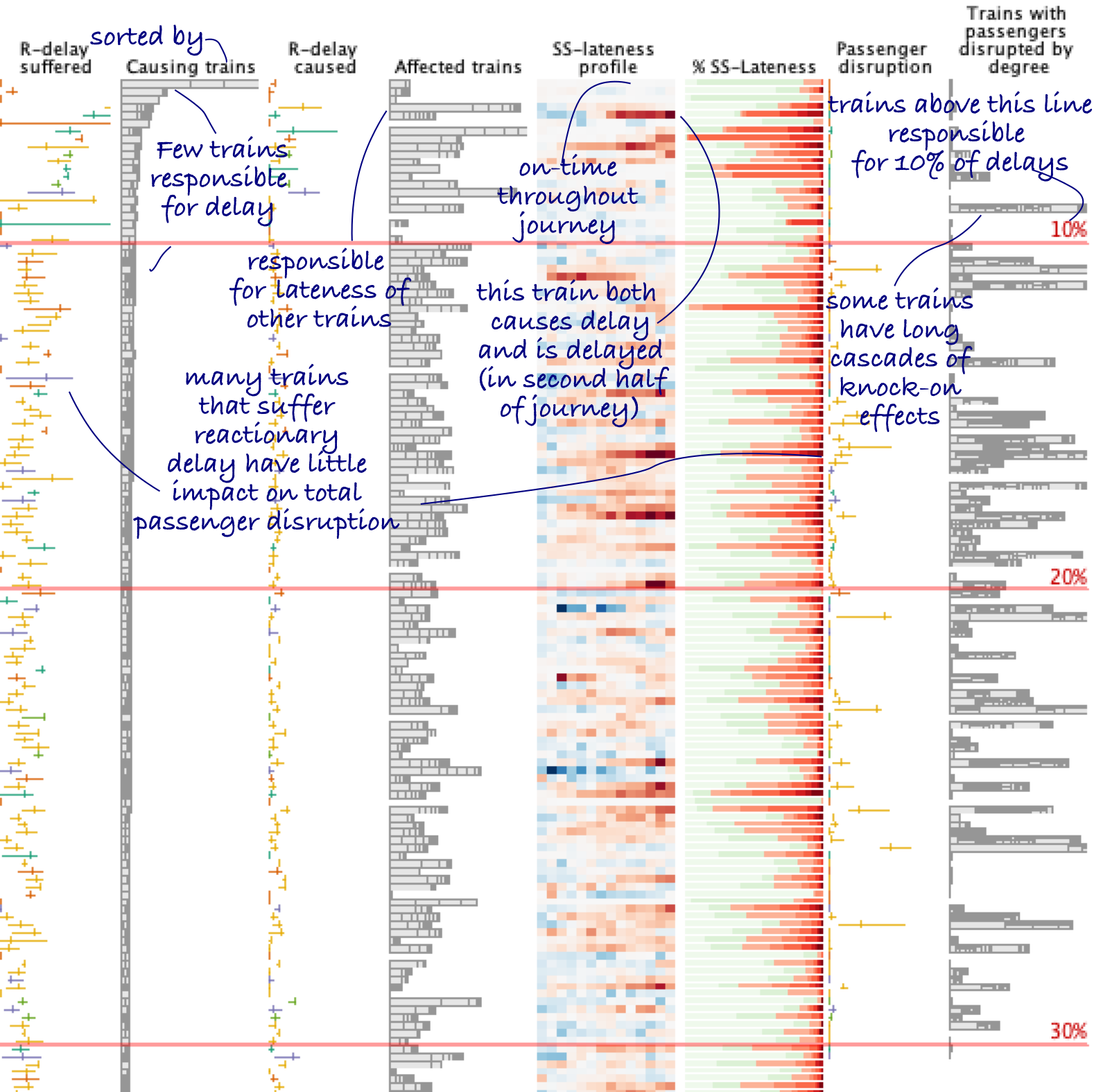}
    \caption{Excerpt zoomed out to the top 20\% of delay-causing trains, sorted by delay caused showing the ``drop-off''. Only a few trains are suffering significant reactionary delay. This is caused by only a few other trains, but they are not causing much reactionary delay to other services and are not themselves delayed much nor causing passenger journey disruption. Trains suffering 10-20\% of the total reactionary delay go on to cause delays to other services.}
    \label{fig:zoomedout}
\end{figure}

R3 and R4 call for a solution that succinctly summarises both the metric value per train (R3) and also its distribution across model runs (R4) for that train. Our solution uses mini-charts with two levels-of-detail variants that correspond to these options representing two different abstractions \cite{beecham2016faceted}. R3 also calls for identifying the most problematic trains and considering the distribution of metric values across all trains. Our solution is to sort trains based on the median or variance values of the specific metric. This helps identify problematic trains in terms of different metrics and the distribution trains (Fig.\ \ref{fig:zoomedout}). Red horizontal decile lines delineate the cumulative delay causes/suffered by trains in 10\% chunks (Fig.\ \ref{fig:zoomedout}). Tooltips provide the numbers where required (e.g.\ Fig.\ \ref{fig:highlighted}).

To summarise the technique: in Zoomable Level-of-Detail ChartTables, rows are \textit{trains}, columns are \textit{composite metrics} that quantify aspects of reactionary delay, visually-represented as \textit{mini-charts} within cells that represent the metric for each train across all model runs. Four types of mini-chart (section \ref{sec:types}) for different metric types have visual representations at two levels-of-detail depending on zoom level (Fig.\ \ref{fig:types}). This semantic zoom (variant of zoom that is not purely geometric \cite{bederson1996pad++}) applies only on the \textit{y}-axis, changing the height of rows but leaving their widths intact.  Where rows are narrow and numerous, low level-of-detail mini-charts summarise across all model runs. Where zoomed rows are wide/tall, high level-of-detail mini-chart depict the details of the whole distribution of model runs within that train. See the \href{https://osf.io/u2ykd/}{video}.

This is a similar approach to Rao and Card's ``Table Lens'' \cite{rao1994table} in which they demonstrate single value metrics where bar (charts) are used for low level-of-detail compact summaries, directly expressed numbers for high level-of-detail, with level-of-detail zooming that applies to subsets of rows and/or column to give focus+context. Our use differs in three ways: (a)~we have more complex metrics; (b)~low and high levels-of-detail are used to summarise across model runs and to show distributions between model runs respectively (instead of charts vs numbers); and (c)~and we apply the level-of-detail zooming on rows only (instead of rows and column) across the whole table (instead of subsets of rows). Our emphasis is their use for helping interpreting probabilistic model outputs.

\begin{figure}[t]
    \centering
    \includegraphics[width=\linewidth]{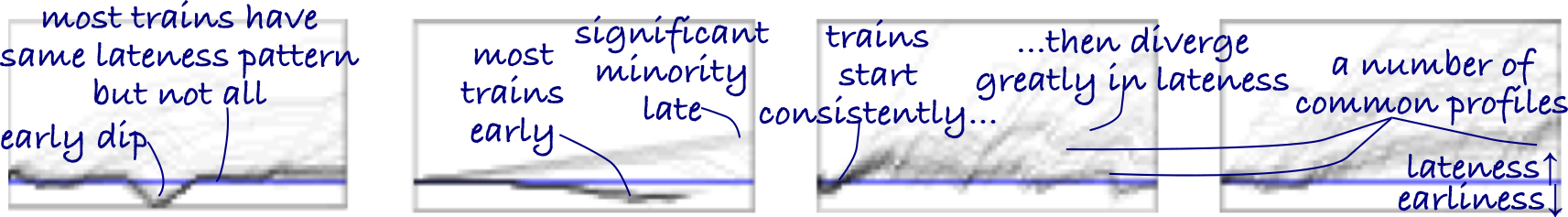}
    \caption{Four examples of high LoD ``LatenessTProfileCharts'' (Fig. \ref{fig:types}b) showing different lateness consistency between model runs.}
    \label{fig:detailed_profiles}
\end{figure}

\subsection{Mini-chart designs}\label{sec:types}

Our metrics (section \ref{sec:supplement_inst}) fall into four types, each with a different graphical representation for both level-of-details, as shown in Fig.\ \ref{fig:types}:

\textbf{(a) LatenessChart} (Fig.\ \ref{fig:types}a) is used where there is a single value per model run; e.g.\ reactionary delay minutes caused. The \textit{low LoD} variant summarises this across the model runs as marks depicting the median and standard deviation. The \textit{high LoD} variant is a dotplot, with a dot for each model runs. Animated jitter \cite{ellis2007taxonomy} reduces misleading effects of overplotting.

\textbf{(b) LatenessTProfileChart} (Fig.\ \ref{fig:types}b) depicts the lateness of a train over the course of its journey. The \textit{low LoD} variant is a temporally binned heatmap coloured by average lateness throughout its journey with a diverging blue/red colour scheme for early/late station stop lateness. In Fig.\ \ref{fig:types}b, the first (at the top) train is on-time throughout its whole journey, whereas the second train starts off late and does not recover after a sharp increase in lateness early in its journey, perhaps indicating a problematic station. The \textbf{high LoD} variant is a line graph where each line represents a model run. In Fig.\ \ref{fig:types}b, the on-time nature of the first train is reflected in all model runs, but for the second train, the model run delays diverge after the aforementioned sharp increase in lateness. Fig.\ \ref{fig:detailed_profiles} shows some alternative high LoD LatenessTProfileCharts with observations.

\textbf{(c) LatenessFreqChart} (Fig.\ \ref{fig:types}c) shows the average frequency of late station stops within lateness categories (using the UK railway industry's categorisation of ``early'', ``0-1min'', ``1-3min', etc \cite{cp6}; see Fig.\ \ref{fig:types}c). The lateness categories are green for $<$1min late categories and red for $>$1min late categories, with darker colours for later lateness categories. The \textit{low LoD} variant uses proportional bar length to indicate the average number of station stops in each lateness category (the greater the proportion of dark red, the later the train). In the \textit{high LoD} variant, this is rotated vertically and one per model run is displayed side-by-side indicating whether there is consistency across model runs. In the \textit{high LoD} variant in Fig.\ \ref{fig:types}c, the second train is consistently late at station stops whereas the third train is sometimes late and sometimes not.

\textbf{(d) AffectTrainsChart} (Fig.\ \ref{fig:types}d) shows the trains that either \textit{affect} or \textit{are affected} by the row's train and \textbf{how much delay} each train is affected or suffered.  For this description, let us \textit{assume these are the trains causing the delay}. The \textit{low LoD} variant is a stacked barchart where each sub-bar represents a train that causes delay to this row's train, where its length indicates the magnitude of delay this caused to this train. In the \textit{high LoD} variant, this is rotated vertically and there is a bar for each model run. Each sub-bar coloured by the train number/identifier so that \textit{consistent hues} indicate that \textit{exactly the same trains are involved} in the different multiple model runs. The first train in Fig.\ \ref{fig:types}d for the low LoD variant is only significantly delayed by two trains (one much more than the other). However the inconsistent colours in the high LoD variant indicates that these are different trains in different model runs. For the fourth train, the low LoD variant indicates that one one train causes the delay (one bar) and the high LoD variant confirmed that this is the same train in all cases. Such insights help decide which trains to focus on for mitigating actions.

In general, the low LoD variants use the \textit{x}-axis to facilitate comparison \textit{between trains} (rows), and high LoD variants use of the \textit{y}-axis to facilitate comparison \textit{between simulation runs}. In the high LoD variants, model runs are ordered by model run number/identifier so that the y-position of the bar across charts corresponds to the same model run. This is important as it help enable a `bad day' to be identified across charts, especially with interactions that highlight identical model runs over the whole table (section \ref{sec:interactions}). Model runs can also be sorted by the value for the metric in help indicate the numerical distribution between model runs.

\subsection{Interactions}\label{sec:interactions}

\textbf{Semantic zoom.} A key contribution of this paper, this triggers mini-charts to be viewed in different level-of-detail modes -- shown side-by-side in Fig.\ \ref{fig:types} -- depending on a zoom level appropriate for the mini-chart. In Fig.\ \ref{fig:types}, the high LoD variant is zoomed such that only the top four rows remain visible. In Fig.\ \ref{fig:zoomedout}, the ChartTable is zoomed out to trains that cause 20\% of the overall delay, enabling the distribution of a given metric between trains to be assessed.

\begin{figure}[t]
    \centering
    \includegraphics[width=\linewidth]{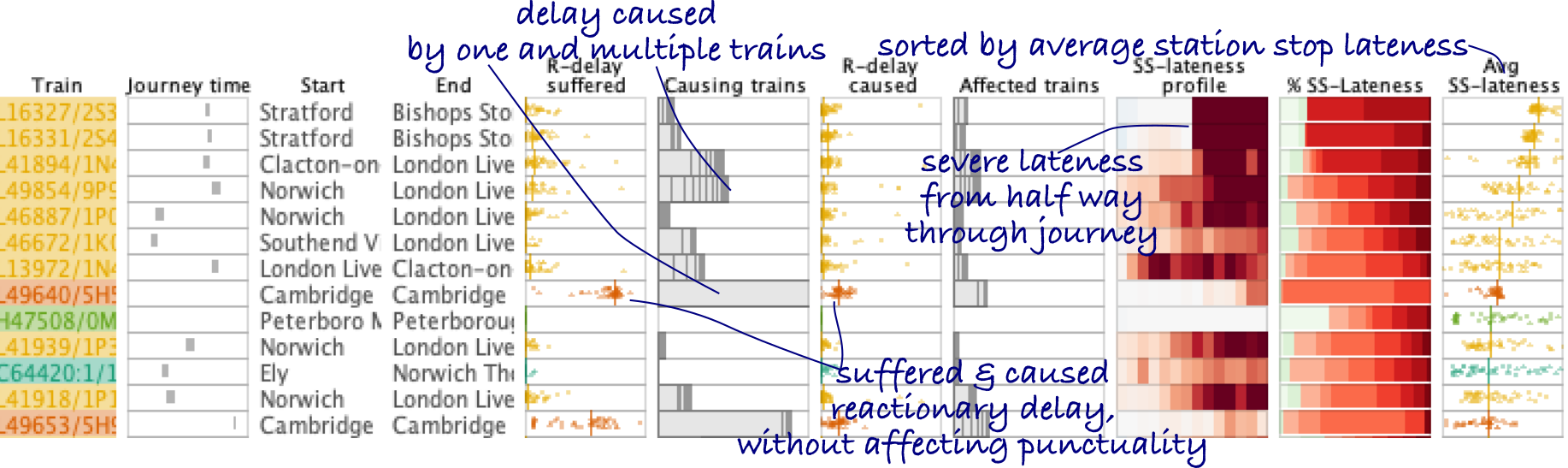}
    \caption{Example sorted by ``average station stop lateness'' with observations made on other aspects of lateness for these trains.}
    \label{fig:excerpt_ss}
\end{figure}

\textbf{Sorting.} Trains can be sorted by metric based on the median or dispersion across model runs. This enables the most or least problematic trains to be identified and the consistency of that metric between the simulation runs and the relationship to other metrics can be determined. See Fig.\ \ref{fig:zoomedout} and Fig.\ \ref{fig:excerpt_ss}.

\textbf{Scaling.} All the cells contain mini-charts. Each metric/column has its own $x$ and/or $y$ axes, which are scaled between zero and the 95th percentile by default. Each column can have its range easily adjusted. The mouseovered cell will always show the entire mini-chart, overlapping its surroundings if necessary.

\begin{figure}[t]
    \centering
    \includegraphics[width=\linewidth]{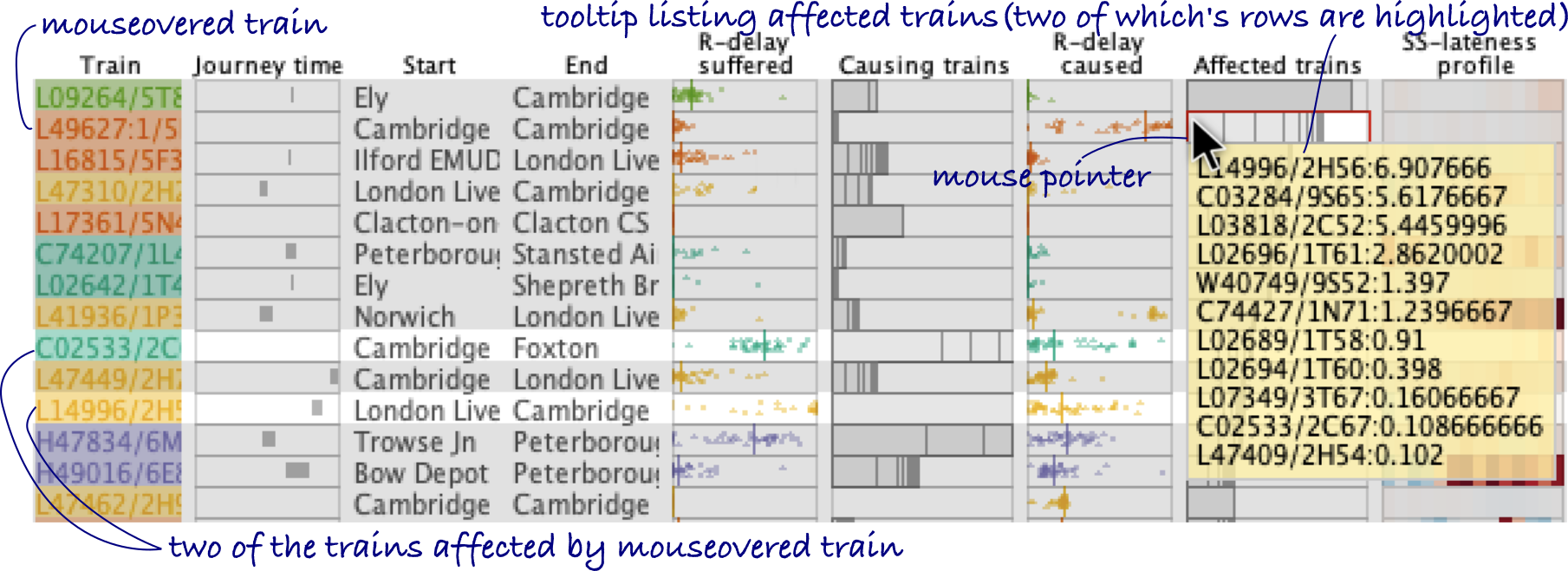}
    \caption{``Highlighting'' interaction that identifies two of the trains involved with the train identified with the mouse cursor.}
    \label{fig:highlighted}
\end{figure}

\textbf{Tooltips} are providing details on demand (example in Fig.\ \ref{fig:highlighted}).

\textbf{Train filtering} allows filtering by train identifier and category.

\textbf{Temporal filtering.} The histogram at the bottom Fig.\ \ref{fig:teaser} shows the number of trains by category throughout the day. These can filtered by time-of-day mouse, preserving sorting and scaling.

\textbf{Highlighting.} Highlighting is an interactive technique for associating data points across the table. Fig.\ \ref{fig:highlighted} illustrates that selecting a train and an ``AffectTrainChart'' metric results all trains/rows in view involved in the metric being highlighted (actually others ``lolighted'') making it easier to access the metric values for this train. The same technique is also used for identifying the same model runs across the table to help assess the impact of ``bad day'' model runs across the dataset.

\section{Experiences, discussion and further work}

\subsection{Reflections on our use of the technique}\label{sec:refl}

\textbf{Alternative `cases'.} The examples presented here have cases (rows) as trains. By summarising the model output by station stops, we have also used \textit{station stops as cases/rows} where the metrics characterise the delays at these locations. This important complementary information enables us to identify problematic locations. For example, identifying key ``pinch-point'' stations at which reactionary delays to many trains lead to huge impacts on other parts of the network gives clues to possible solutions.

\textbf{Adding metrics.} It was easy to add additional metrics as the work progressed. This included adding a set of metrics where delays are weighted passenger numbers by highlighting the trains/stations in terms of impact to passengers.

\textbf{Visual scalability.} We have been able to deal with hundreds of simulation runs and thousands of trains, including the ability to zoom-out to all trains. A limitation of our implementation is the requirement for at least one pixel per train and model run (the latter with high LoD variants). We can improve the visual scalability by appropriately sampling trains and model runs where not enough pixels are available to depict them all. This may need to happen dynamically, depending on zoom level.

\textbf{Sorting.} In our implementation, sorting is limited to simple point/dispersion value to summarise the metric by train. We are implementing sorting based on different points on the numerical distribution. For example, 80th percentiles are sometimes used for identifying plausibly likely ``bad days''.

\textbf{Historical data.} Train lateness data is routinely collected and used to assess past performance. We can use these data in ChartTables, where real days replace model runs and where the distribution of lateness is based on what actually happened. Since only the effects of delays are recorded, there is a lack of detail on the \textit{mechanism of the delays} afforded by our model. However, the Train Operating Companies we worked with found that visual analysis of historical data was valuable, albeit with fewer metrics.

\subsection{Potential uses in other contexts}

\textbf{Interpreting probabilistic model outputs.} Stochastic Monte-Carlo-style ABMs are in widespread use, generating large amounts of data, usually too much to consider in detail \cite{Grainger2016}. In most cases, these can be summarised in a case-by-variable structure. The ``alternative case'' point above illustrates the flexibility of summarising model outputs by different case-types and interpreting them within ChartTables. Since our four metric types are specific to our application, it is likely that additional mini-chart designs will be needed.

\textbf{Other data and other mini-charts.} Using other model outputs or datasets will likely necessitate additional mini-charts. Fortunately, the design space of such mini-charts is enormous \cite{munzner2014visualization} and can be based on existing visualisation idioms (as most of ours were) or customised designs. Whether variables and cases are rows or columns influences the effectiveness of the \textit{x-} and \textit{y-}axes. Where rows are cases, using \textit{x-} axis for low LoD variant facilitates comparison across rows. For high LoD variants, using the \textit{y-}axis facilitates comparison between model-runs within cases.

\textbf{Other levels-of-detail.} Although two LoDs were appropriate for our use, different conceptualisations of LoD will be appropriate in other contexts; for example, corresponding to levels of abstraction \cite{beecham2016faceted}, aggregation and/or hierarchical data \cite{furmanova2020taggle, li2024coinsight,li2022hitailor}.

\subsection{Use in the UK railway industry}

The original purpose of our work with the UK railway industry was to develop and determine the feasibility of an ABM/visualisation approach to understanding reactionary delay. We expected the value to industry to be strategic, such as identifying general guidelines for timetable design. However, we found that its use more operationally was more important to the Train Operating Companies we worked with. They were interested in investigating specific timetable proposals and identifying problematic trains with reasons, and using these to tweak then test these modified timetables.  Although much of the analytical work was been carried out by Risk Solutions, Zoomable LoD ChartTables play an important narrative role when presenting modelling results to clients.

Feedback has been positive. Simon Greenwood (Performance Manager, GWR) finds them \textit{``really helpful, being able to see where reactionary delays were likely to be, and where they were coming from, delivered new insights and helped us see where to focus our attention''} and Marc Ware (Performance Manager, Greater Anglia) stating that \textit{``it helped us work through a very complex system to understand the individual levers we have to use to deliver change'}'. Actionable evidence is also valued, with Mark Walker (Performance Manager, ARL) citing the \textit{``useful evidence to support our discussions with Network Rail about designing a better timetables''} and Marc Ware saying \textit{``this is giving us the evidence we need\ldots and focus on building a more robust timetable''}.

We are now licensing software that incorporates the ABM and visualisation, so some Train Operating Companies will now be using them analytically. Training sessions and the feedback this generates is informing further development.

\section{Conclusion}

Studying reactionary delay in trains is a good example of how visual analytics can help analysts consider nuance and variation in data that is normally averaged out. Tabular layouts are a simple, familiar and effective means to represent variables and cases. Mini-charts provides a huge range of possibilities for representing multivariate composite metrics and data distributions, including qualifications of uncertainly. Zoom-based level-of-detail variants of these mini-charts enabled us to consider distributions of summaries by case and and distributions within cases of Monte-Carlo-style stochastic model outputs. There is plenty of scope to generalise to other application areas.

\section{Supplemental Material Instructions}
\label{sec:supplement_inst}

Supplementary materials are available at \url{https://osf.io/u2ykd/} including \textit{a video} and \textit{a list of our train metrics}. The computer code is not available as portions of it are owned commercially, however it is relatively easy to implement and the authors are happy provide further information.

\acknowledgments{
The authors wish to thank the \href{https://www.rssb.co.uk/}{Rail Safety and Standards Board (RSSB)}, \href{https://www.networkrail.co.uk/}{Network Rail}, \href{https://www.gwr.com/}{Great Western Railways}, \href{https://www.greateranglia.co.uk/}{Greater Anglia}, the \href{https://www.westcoastpartnershipdevelopment.co.uk/}{West Coast Partnership Development}, \href{https://www.arrivaraillondon.co.uk/}{Arriva Rail London} and \href{https://web.archive.org/web/20221229005042/https://www.crossrail.co.uk/}{CrossRail} for their funding and other support with this work.}

\bibliographystyle{abbrv-doi}

\bibliography{main}

\begin{thebibliography}{10}

\bibitem{bederson1996pad++}
B.~B. Bederson, J.~D. Hollan, K.~Perlin, J.~Meyer, D.~Bacon, and G.~Furnas.
\newblock Pad++: A zoomable graphical sketchpad for exploring alternate
  interface physics.
\newblock {\em Journal of Visual Languages \& Computing}, 7(1):3--32, 1996.

\bibitem{beecham2016faceted}
R.~Beecham, C.~Rooney, S.~Meier, J.~Dykes, A.~Slingsby, C.~Turkay, J.~Wood, and
  B.~W. Wong.
\newblock Faceted views of varying emphasis (favves): a framework for
  visualising multi-perspective small multiples.
\newblock In {\em Computer Graphics Forum}, vol.~35, pp. 241--249. Wiley Online
  Library, 2016.

\bibitem{bertin1983semiology}
J.~Bertin.
\newblock {\em Semiology of graphics}.
\newblock University of Wisconsin press, 1983.

\bibitem{ellis2007taxonomy}
G.~Ellis and A.~Dix.
\newblock A taxonomy of clutter reduction for information visualisation.
\newblock {\em IEEE transactions on visualization and computer graphics},
  13(6):1216--1223, 2007.

\bibitem{furmanova2020taggle}
K.~Furmanova, S.~Gratzl, H.~Stitz, T.~Zichner, M.~Jaresova, A.~Lex, and
  M.~Streit.
\newblock Taggle: Combining overview and details in tabular data
  visualizations.
\newblock {\em Information Visualization}, 19(2):114--136, 2020.

\bibitem{Grainger2016}
S.~Grainger, F.~Mao, and W.~Buytaert.
\newblock {Environmental Modelling \& Software Environmental data visualisation
  for non-scientific contexts: Literature review and design framework}.
\newblock {\em Environmental Modelling and Software}, 85:299--318, 2016. doi:
  {{%
10\hspace{.1pt}\discretionary{.}{%
}{.}\hspace{.4pt}1016\discretionary{/}{%
}{/}j\hspace{.1pt}\discretionary{.}{%
}{.}\hspace{.4pt}envsoft\hspace{.1pt}\discretionary{.}{%
}{.}\hspace{.4pt}2016\hspace{.1pt}\discretionary{.}{%
}{.}\hspace{.4pt}09\hspace{.1pt}\discretionary{.}{%
}{.}\hspace{.4pt}004}}


\bibitem{li2024coinsight}
G.~Li, R.~Li, Y.~Feng, Y.~Zhang, Y.~Luo, and C.~H. Liu.
\newblock Coinsight: Visual storytelling for hierarchical tables with connected
  insights.
\newblock {\em IEEE Transactions on Visualization and Computer Graphics}, 2024.

\bibitem{li2022hitailor}
G.~Li, R.~Li, Z.~Wang, C.~H. Liu, M.~Lu, and G.~Wang.
\newblock Hitailor: Interactive transformation and visualization for
  hierarchical tabular data.
\newblock {\em IEEE Transactions on Visualization and Computer Graphics},
  29(1):139--148, 2022.

\bibitem{munzner2014visualization}
T.~Munzner.
\newblock {\em Visualization analysis and design}.
\newblock CRC press, 2014.

\bibitem{reactionary}
{National Rail}.
\newblock How a delay to services in one area can affect trains elsewhere in
  the country.
\newblock Technical report, 2019.

\bibitem{cp6}
{Network Rail}.
\newblock From monday 1 april 2019, the uk rail industry introduced a new, more
  detailed and precise set of measures to better understand how timely our
  trains are – from station to station, and from minute to minute.
\newblock Technical report, 2024.

\bibitem{perin2014revisiting}
C.~Perin, P.~Dragicevic, and J.-D. Fekete.
\newblock Revisiting bertin matrices: New interactions for crafting tabular
  visualizations.
\newblock {\em IEEE transactions on visualization and computer graphics},
  20(12):2082--2091, 2014.

\bibitem{rssb}
{Rail Safety and Standards Board (RSSB)}.
\newblock Call for research: Data sandbox: Improving network performance.
\newblock Technical report, 2017.

\bibitem{railsback2019agent}
S.~F. Railsback and V.~Grimm.
\newblock {\em Agent-based and individual-based modeling: a practical
  introduction}.
\newblock Princeton university press, 2011.

\bibitem{rao1994table}
R.~Rao and S.~K. Card.
\newblock The table lens: merging graphical and symbolic representations in an
  interactive focus+ context visualization for tabular information.
\newblock In {\em Proceedings of the SIGCHI conference on Human factors in
  computing systems}, pp. 318--322, 1994.

\bibitem{slingsby2018tilemaps}
A.~Slingsby.
\newblock Tilemaps for summarising multivariate geographical variation.
\newblock {\em Workshop on Visual Summarization and Report Generation at VIS
  2018}, 2018.

\bibitem{slingsby2023gridded}
A.~Slingsby, R.~Reeve, and C.~Harris.
\newblock Gridded glyphmaps for supporting spatial covid-19 modelling.
\newblock In {\em 2023 IEEE Visualization and Visual Analytics (VIS)}, pp.
  1--5. IEEE, 2023.

\bibitem{wickham2012glyph}
H.~Wickham, H.~Hofmann, C.~Wickham, and D.~Cook.
\newblock Glyph-maps for visually exploring temporal patterns in climate data
  and models.
\newblock {\em Environmetrics}, 23(5):382--393, 2012.

\end{thebibliography}
\end{document}